\documentstyle[12pt]{article}
\def\bib#1{$^{\ref{#1}}$}

\begin{document}

\centerline {\bf Decay of cosmological constant as Bose condensate 
evaporation} 
\vskip0.2in
\centerline{\bf Irina Dymnikova}
\centerline{\it Institute of Mathematics and Informatics,}
\centerline{\it University of Warmia and Mazury in Olsztyn}
\centerline{\it Zolnierska 14, 10-561 Olsztyn, Poland; 
e-mail: irina@matman.uwm.edu.pl}
\centerline{\bf Maxim Khlopov}
\centerline{\it Center for Cosmoparticle Physics "Cosmion", 125047,
Moscow, Russia; e-mail: mkhlopov@orc.ru}
\vskip0.4in
{\bf Abstract}
\vskip0.2in

We consider the process of decay of symmetric vacuum state as evaporation 
of a Bose condensate of physical Higgs particles, 
defined over asymmetric vacuum state. Energy density of their selfinteraction
is identified with cosmological constant $\Lambda$ in the Einstein equation.
$\Lambda$ decay then provides dynamical realization of spontaneous
symmetry breaking.
The effective mechanism is found for damping of 
coherent oscillations of a scalar field,
leading to slow evaporation regime as
the effective mechanism for $\Lambda$ decay responsible for
inflation without special
fine-tuning of the microphysical parameters. 
This mechanism is able to incorporate reheating, generation 
of proper primordial
fluctuations, and nonzero cosmological constant today.
\vskip0.4in
Any thing which contributes to the stress-energy tensor as $T_{\mu\nu}
=\rho_{vac}g_{\mu\nu}$, behaves like a cosmological 
term $\Lambda g_{\mu\nu}$ in the Einstein equation.
Developments in particles and quantum field theory, as well as
confrontation of models with observations in cosmology, compellingly
indicate that the cosmological constant $\Lambda$ ought
to be treated as a dynamical quantity (for recent review see 
\bib{cooperstock}). 
At the very early stage of the Universe evolution huge value of
cosmological constant is needed corresponding to $\rho_{vac}\geq {\rho}_{GUT}$,
to drive inflation 
as providing the reason for the expansion of the Universe 
and its isotropy and large scale 
homogeneity \bib{gliner}$^{,}$\bib{us75}$^{,}$\bib{guth}$^{,}$\bib{me86}
$^{,}$\bib{linde}$^{,}$\bib{olive}.
Now cosmological constant is estimated from the variety of
observational data at the level $\rho_{vac}\sim{(0.6-0.8)\rho_{today}}$
\bib{neta}$^{,}$\bib{ostriker}.

Several mechanisms have been proposed involving and supporting
a negative vacuum energy density growing with time to cancel
initial pre-existing positive cosmological constant $\Lambda$.
All those mechanisms utilize the basic
property of de Sitter space-time - its quantum and semiclassical
instabilities \bib{star}$^{,}$\bib{myhrvold}$^{,}$\bib{ford}$^{,}$
\bib{dolgov}$^{,}$\bib{dolg}$^{,}$\bib{mottola}$^{,}$\bib{hiscock}.

In scalar field dynamics the potential of
a scalar field $V(\phi)$ plays the role of an effective cosmological
constant in regimes when its derivatives
are close to zero. Starting from such a regime as an initial state, 
field equations typically lead to successive coherent field 
oscillations \bib{kolb}$^{,}$\bib{olive}.
In the context of inflationary models with effective $\Lambda$ related
to inflaton field in slow rolling regime, the further decay
of coherent oscillations involves inflaton interactions with other
fields in models of 
preheating \bib{kof}$^{,}$\bib{rob}$^{,}$
\bib{boyan}$^{,}$\bib{zla}.

On the other hand, we can treat the classical scalar field 
as the Bose-Einstein condensate of physical quanta of this field 
defined over its ground state (true vacuum). 
In our previous papers \bib{us95}$^{,}$\bib{us96} it was shown 
that process of emergence 
of massive scalar particles in (from) de Sitter
vacuum looks like evaporation of a Bose condensate.
In our paper \bib{us98} we proposed the model of self-consistent 
inflation in which the same self-interacting scalar field is responsible 
for both initial value of $\Lambda$ and its further decay.

In the case of the Higgs field this approach  involves
both space-time and particle internal symmetries. 
In gauge theories  
mechanisms of spontaneous symmetry breaking 
imply that 
unbroken symmetry state is false vacuum state.
At the same time this is the highly symmetric state 
of space-time geometry invariant under de Sitter group. 
In Ref.\bib{us98} we interpreted this state 
as Bose-Einstein condensate 
of physical Higgs particles whose self-interaction energy density 
corresponds to a scalar field potential in the state with
the unbroken symmetry. 
The process of decay
of inflationary vacuum appears then as slow evaporation of a Bose
condensate  responsible for inflation and further
transition to the standard FRW model without
special fine-tuning in initial conditions for inflation.
Dynamics of the cosmological term is directly related to the 
hierarchy of particle symmetries breakings, which makes such an
approach physically self-consistent \bib{us98}.

In the present paper we identify cosmological constant $~ \Lambda~ $ 
with the energy density of self-interaction of scalar bosons bound
in the condensate.
We show the existence of the slow evaporation mechanism
which seems to be generic for dynamics of Bose condensates. 
This mechanism for $\Lambda$ decay 
produces an effective kinetic damping  
involving decoherence of coherent scalar field oscillations 
due to self-interaction and back reaction of decay products.

\vskip0.1in
Let us start with the simplest example of the
Higgs field $\phi$
$$V(\phi)=\frac{\lambda}{4}({\phi}^2-{\phi}_0^2)^2\eqno(1)$$
In the context of particle theories, constant part of the Higgs
potential is usually omitted \bib{quigg}. However, in the cosmological
context this term  becomes important. 
In theories with spontaneous symmetry breaking the vacuum expectation
value of the Higgs field, which couples to bosons and fermions to give them 
masses, plays the role of an order parameter. The nonzero vacuum
expectation value in the asymmetric vacuum state $<\phi>=\phi_0$,
is interpreted as the development of a condensate of $\phi$ particles,
leading to the spontaneous symmetry breaking \bib{kolb}$^{,}$\bib{quigg}. 

On the other
hand, $\phi$ particles are not physical particles, they are tachionic,
with the imaginary mass, that reflects the instability of the symmetric
state of the theory. In the asymmetric, physically stable vacuum state
(the state with the broken symmetry) physical particles are $~ \chi~ $
particles, related to $\phi$ particles by $\phi=\chi+\phi_0$.
It is $\chi$ particles who acquire the mass by the Higgs mechanism,
and whose vacuum state is the true vacuum with zero potential and with
zero expectation value, $<\chi>=0$. Therefore, in terms of particles 
$\chi$, the true vacuum of theories with spontaneous symmetry breaking,
cannot be a condensate, and we would rather have to treat 
the symmetric vacuum state of a theory as a condensate of $\chi$
particles in bound state. Replacing $\phi$ condensate by $\chi$
condensate sheds some light on the origin of $\Lambda$, which in
$\phi$ condensate picture, where an effective $\Lambda$ is related to the
state of zero field \bib{guth}, looks mysterious. 
As we shall see below, in $\chi$
condensate picture effective $ \Lambda$ is related to nonzero value of the
field $\chi$ as its energy density in symmetric state.

The Higgs field $\chi$ in this simple model is described by the
Lagrangian \bib{us98}
$${\cal L}=\sqrt{-g}\biggl[\frac{1}{2}g^{\mu\nu}{\chi}_{;\mu}{\chi}_{;\nu}
-\frac{\lambda}{4}(2{\phi}_0\chi+{\chi}^2)^2\biggr]\eqno(2)$$
where we dropped, for simplicity, the indices of internal variables.
The potential takes the form
$$V(\chi)=\lambda{\phi}_0^2{\chi}^2+\lambda{\phi}_0{\chi}^3
+\frac{\lambda}{4}{\chi}^4\eqno(3)$$
In the state with unbroken symmetry
$~<\chi>=-{\phi}_0~$, $~<{\chi}^2>={\phi}_0^2~$, 
$~<{\chi}^3>=-{\phi}_0^3~$, and the energy density of a condensate
of $~\chi~$ particles is given by
$$<V(\chi)>=\frac{\lambda}{4}{\phi}_0^4\eqno(4)$$
We see that the constant term, playing the role of the cosmological
constant $\Lambda$ in the Einstein equation, is identified with the
energy density of condensate of $~\chi~$ particles being
precisely the energy density of their self-interaction.

The potential (3)
describes physical particles $\chi$ with masses of Higgs bosons
$~m=\sqrt{2\lambda}{\phi}_0$, self-interacting and interacting
with condensate. The term $~\lambda{\phi}_0{\chi}^3~$
corresponds to decay of a condensate via the channel
$~{\phi}_0\rightarrow{3\chi}~$, and describes evaporation
of Higgs bosons from condensate of   $~\chi~$ particles.
The term $~\lambda{\chi}^4/4~$ reproduces the runaway
particle production discovered by Myhrvold \bib{myhrvold}.
The difference from the Myhrvold result is in the origin of $~\Lambda$.
Gravity-mediated decay of $~\Lambda~$ in Myhrvold approach is due to
particle creation by gravitational field generated by pre-existing
$~\Lambda~$ not related to created particles.
In our approach $\Lambda\sim{m^2{\phi}_0^2}$
 is the energy density of self-interaction of the same
particles bound within a condensate.

The Hubble parameter $~H~$ during the $~\Lambda~$ dominated stage is
$H\sim{\sqrt{\Lambda}}$. In the context of particle theories with
${\phi}_0\ll m_{Pl}$, our case corresponds to creation of bosons
with $m\gg H$.
Therefore our mechanism differs from that proposed by Mottola 
who studied creation
of particles  by gravitational field
via the Hawking quantum evaporation which leads to the exponential
suppression of masses $~m\gg H$ \bib{mottola}. 
In the Mottola mechanism light scalar particles with $~ m < H~ $ 
are evaporated from de Sitter horizon induced by pre-existing $~ \Lambda$.
 In our mechanism Higgs bosons with $~m\gg H~$ are evaporated
from the bound state within a condensate into the free states.

Our approach differs also from Parker and Zhang theory of relativistic
charged condensate as a source of slow rolling inflation \bib{parker}. 
In our aproach condensate of $~\chi~$
particles is essentially globally neutral, since it corresponds to
the state with unbroken symmetry - totally symmetric state in both
space-time and internal degrees of freedom \bib{us98}. 
The condensate decays by
evaporation, as well as by runaway production of $~\chi~$ particles
which corresponds to conversion of energy of initial globally neutral 
state into thermal energy of $~\chi~$ particles.

Roughly, scenario of $~ \Lambda~$ decay looks as follows. 
Within a $\chi$ condensate $\chi$ particles have four-momenta 
$k\rightarrow 0$. Indeed, the Klein-Gordon equation for 
the potential (3) in the condensate regime 
($\chi=-\phi_0; V^{\prime}(\chi)=0$) reduces to 
$\ddot{\chi}+3H\dot{\chi}+\Gamma\dot{\chi}=0$,
where $\Gamma$ is a decay rate. Its solution
reads $\chi=-\phi_0-\phi_0 e^{-(3H+\Gamma)t}$. The time-dependent
mode decays rapidly for any $\Gamma$ stabilizing particle state
in condensate with zero four-momentum ($\dot{\chi}\rightarrow 0$).
This means that the $\chi$ condensate is the classical collective 
state within which decay of $~ \chi~ $ particles is impossible.
(Their de Broglie wavelength far exceeds the horizon, so that
the asymptotic states cannot be defined for the quantum transitions 
corresponding to decay inside a condensate.) 
Therefore decay of $~ \chi~ $ particles occurs in free states
which correspond to the coherent field oscillations.

The  fluctuations
$~\delta=\chi+{\phi}_0~$ over the state with unbroken symmetry
$~<\chi>=-{\phi}_0~$ are described by the potential
$$V(\delta)=\frac{\lambda}{4}{\phi}_0^4
-\frac{\lambda}{2}{\phi}_0^2{\delta}^2
+\frac{\lambda}{4}{\delta}^4\eqno(5)$$
which reflects the instability of the state with
unbroken symmetry. Wrong sign of the mass term corresponds to
the excitation of the growing mode of perturbation $~\delta~$
over the symmetric vacuum state, leading to its decay.
The evolution of fluctuations is governed by the Klein-Gordon equation
$$\ddot{\delta}+3H\dot{\delta}-\frac{m^2}{2}
\delta+\lambda{\delta}^3=0\eqno(6)$$
Let us estimate the characteristic time scale for the linear stage
of development of instability, neglecting the ${\delta}^3$ term and
taking into account that in our case ($~ \phi_0\ll m_{Pl}~ $)
$$\frac{m}{H}=\sqrt{\frac{3}{\pi}}\frac{m_{Pl}}{{\phi}_0}\gg 1\eqno(7)$$
In this approximation the solution to the Eq.(6) is given by
$$\delta(t)=C_1 e^{-\frac{m}{\sqrt{2}}t}+C_2 e^{\frac{m}{\sqrt{2}}t}\eqno(8)$$
Growing mode of $~\delta(t)~$ corresponds to decay of condensate with
characteristic timescale $~\tau\sim{m^{-1}}~$.
To estimate efficiency of decay we introduce the fluctuation density
$$n_{\delta}=\frac{m}{4}{\delta}^2\eqno(9)$$
Then
$$V(\delta)=\frac{\lambda}{4}{\phi}_0^4-mn_{\delta}
+\frac{4\lambda{n_{\delta}}^2}{m^2}\eqno(10)$$
The last term can be interpreted as self-interaction of fluctuations
with the reaction rate $~4\lambda/m^2~$ leading to effective decoherence 
of coherent field oscillations.
In the minimum of the potential (10) fluctuation density 
$~n_{\phi}=m^3/8\lambda~$ corresponds, by Eq.(9), 
to $~{\delta}^2={\phi}_0^2~$. We see that stationary
distribution of fluctuations around $~\delta={\phi}_0~$ 
looks like a gas of $~\chi~$ particles with masses 
$~m=\sqrt{2\lambda}{\phi}_0$,
evaporated from the state within a condensate into the 
fluctuations.
The potential (10) achieves its
minimum $~V(\delta)=0~$, which corresponds to the total decay 
of $~\Lambda~$ condensate into decoherent gas of $~\chi~$ particles,
in characteristic time $~\tau\sim{m^{-1}}\ll{H^{-1}}$. 
So, the decay of $\chi$ condensate provides the efficient 
mechanism for $~\Lambda~$ decay.
 
We see, that identifying $~V(-\phi_0~) $ with the energy density 
of self-interaction of physical particles $ ~\chi~$ in the bound state 
of a Bose condensate,
we can treat instability of de Sitter vacuum as dynamical
 realization of symmetry breaking.
 The process of
$~ \Lambda~ $ decay proceeds in this picture 
as the Bose condensate evaporation.
\vskip0.1in
Now let us show that the effective mechanism exists related
to this approach which makes the process of evaporation slow enough
to guarantee $e-$folding needed for inflation.

Higgs bosons are generally unstable. 
Both $~ \chi~ $ particles
and products of their decay interact with
Hawking radiation from de Sitter horizon
and with each other. This leads not only to decoherence of $~ \chi~ $ particles
but also to appearance of 
relativistic particles over the condensate with the
effective equation of state $~p=\varepsilon/3~$. Presence of relativistic
gas in thermodynamic equilibrium with condensate facilitates its further
decay. 

The dynamical role of not exponentially small thermal component
in the evolution of an inflaton field has been studied by Berera
and Li-Zhi Fang \bib{berera} and then incorporated into the scenario
called warm inflation \bib{warm}. The classical equation of motion
for a scalar field $\varphi$ in the de Sitter universe reads \bib{kolb}
$$\ddot{\varphi}+3H\dot{\varphi}+\Gamma_{\varphi}\dot{\varphi}
+V^{\prime}(\varphi)=0$$
The friction term $\Gamma_{\varphi}$ is introduced phenomenologically
to describe the decay of the inflaton field $\varphi$ due to its
interaction with thermal component. Berera and Li-Zhi Fang have shown, 
first, that
if $\Gamma_{\varphi}\sim{H}$, then thermal component can play essential
role in generation of a primordial density fluctuations \bib{berera}.
Second, in the regime $\Gamma_{\varphi}\gg H$ thermal component
becomes dominant in the equation of motion leading to warm
inflation \bib{warm}.
The question of self-consistency of this regime as governed by thermal
components has been addressed in the 
papers \bib{warm}$^{,}$\bib{crit}$^{,}$\bib{bel}.

In our case the mechanism leading to this condition which we 
write as $\Gamma_{\chi}\gg H$ is not related to thermal components.
It is related to the effective friction due to interaction of free particles
with the ensemble of $\chi$ particles bound in condensate with
the large occupation number, which damps oscillations and leads
to stationary regime of slow evaporation of $\chi$ condensate.

 Let us show that decoherence of $~ \chi~ $ particle states and back reaction
of their relativistic decay products lead to the effective damping
of fluctuations.
The kinetic equations can be written in the standard way (see, e.g.\bib{kolb}).
The kinetic equation describing the growth and decay of fluctuations is
$$\frac{dn_{\chi}}{dt}=mn_{\chi}-\Gamma n_{\chi}-n_rn_{\chi}{\sigma}
-3Hn_{\chi}\eqno(11)$$
where $\sigma$ is the cross-section of the interaction of $\chi$
particles with relativistic products of their decay. In the units
$\hbar =c=1$ the reaction rate in the kinetic equations coincides
with $\sigma$.
The first term in the right hand side describes creation of $~\chi~$
particles, the second - their decay, the third - their interaction
with products of decay, and the fourth - their redshifting.

The kinetic equation for products of decay, effectively relativistic
matter with the equation of state $~p=\varepsilon/3~$, reads
$$\frac{dn_r}{dt}=-3Hn_r+n_rn_{\chi}{\sigma}+\Gamma n_{\chi}\eqno(12)$$

In the equilibrium
$$mn_{\chi}-\Gamma n_{\chi}-n_rn_{\chi}{\sigma}
-3Hn_{\chi}=0$$
$$ -3Hn_r+n_rn_{\chi}{\sigma}+\Gamma n_{\chi}=0\eqno(13)$$
Decay of $~\chi~$ particles
into light species implies  $~m\gg \Gamma~$, which corresponds 
to applicability of
perturbation theory for calculations of decay,
and is valid in models with coupling less than the unity.
Then, taking into account Eq.(7), we get
$${\rho}_r=\frac{m^2}{{\sigma}}; ~~~~~~~
{\rho}_{\chi}=\frac{3mH}{{\sigma}}\eqno(14)$$
Equilibrium density of relativistic particles $\rho_r$
is achieved when the density of evaporated and decayed 
$\chi$ particles 
$$\rho_{\chi d}\sim{\Gamma\frac{1}{m}\rho_{vac}}=
\Gamma\frac{1}{m}\frac{m^2 \phi_0^2}{8}$$ 
satisfies the condition
$$\Gamma\frac{1}{m}\frac{m^2 \phi_0^2}{8} > \frac{m^2}{\sigma}$$
This gives the lower limit on the characteristic width of 
$\chi$ particles decay
$$\Gamma > \frac{8m}{\phi_0^2 {\sigma}}\eqno(15)$$
If this condition is satisfied,
the potential evolves not to the value $~ V(\delta)=0$,
but to the value $~ V_{max}-\rho_{\chi}-\rho_r$, becoming successively more
flat. 
Slow evaporation of $~ \chi~ $ condensate acts in such a way 
to flatten the potential
near its symmetric state. 

We see that back reaction of evaporated particles and products
of their decay produces an effective damping of scalar field
oscillations which leads to effective flattening of
initially nonflat potential and provides mechanism responsible for
inflation without fine-tuning of the potential parameters.
This is qualitatively similar to effective flattening found
around spinoidal line in the Hartree-truncated theory of
spinoidal inflation \bib{spin} (which involves fine-tuning
at the slow rolling stage preceeding the spinoidal regime). 

The process of $~\Lambda~$ decay is governed by the equation
$$\frac{d{\rho}_{vac}}{dt}=-3H({\rho}_{\chi}+{\rho}_r
+\frac{1}{3}{\rho}_r)=-4H\frac{m^2}{{\sigma}}\eqno(16)$$
We can estimate the characteristic time of decay for
two limiting cases of minimal and maximal cross-section $~{\sigma}~$.
The lower limit on cross-section $~ \sigma~ $ is  
$~{\sigma}=4\pi/{m^2}~$ which is
the hard ball approximation cross-section for scattering of particles 
of masses $~m/2$. In this case
$$\frac{d{\rho}_{vac}}{dt}=-\frac{H}{\pi}m^4\eqno(17)$$
The law for $~\Lambda~$ decay 
$${\rho}_{vac}={\rho}_0\biggl(1-\frac{t}{\tau}\biggr)^2; ~~~ 
\tau=\sqrt{\frac{3\pi}{2}}\frac{m_{Pl}\sqrt{\rho_0}}{m^4}
=\sqrt{\frac{3\pi}{32\lambda}}\frac{m_{Pl}}{m^2}\eqno(18)$$
gives the e-folding number 
$$H\tau=\frac{\pi}{8}\frac{1}{\lambda}\eqno(19)$$
and sufficient inflation for reasonable values of coupling $\lambda$.
The characteristic time for reheating is $\tau\sim{{\lambda H}^{-1}}$
and the reheat temperature $T_{RH}\sim{{\lambda}^{1/4}m}$.

The upper bound for $~{\sigma}~$ is given by 
${\sigma}=\pi/{H^2}$.
In this case
$$\frac{d{\rho}_{vac}}{dt}=-\frac{4}{\pi}H^3m^2;~ ~ ~ 
{\rho}_{vac}=\frac{{\rho}_0}{(1+t/{\tau})^2}\eqno(20)$$
where
$$\tau=\biggl(\frac{3^3}{2^{11}{\pi}}\biggr)^{1/2}\frac{1}{{\lambda}^{3/2}}
\biggl(\frac{m_{Pl}}{{\phi}_0}\biggr)^4 {\tau}_{Pl}\eqno(21)$$
The e-folding number is then
$$H\tau=\frac{3}{32}\frac{1}{\lambda}
\biggl(\frac{m_{Pl}}{{\phi}_0}\biggr)^2\eqno(22)$$
and, for the considered case $~{\phi}_0\ll m_{Pl}$,
 inflation is sufficient for any $~\lambda$.
Reheating temperature is $~T_{RH}\sim{{\lambda}^{1/4}H}$.

More detailed investigation of dynamics of a vacuum decay needs
particular model for calculating ${\sigma}$, but the results
will be within the range between the cases of minimal and maximal
${\sigma}$. For example, the picture of evaporation investigated
in the Ref.\bib{us95} corresponds to the case of evaporation
of Higgs bosons and their reheating to the Hawking temperature
$T_{RH}\sim{H}$ \bib{hawking}. The rate of $\Lambda$ decay is given
in this case by
$$\frac{d{\rho}_{vac}}{dt}=-3Hm(mH)^{3/2}\eqno(23)$$

Here we considered $\Lambda$ decay in the case
of vacuum dominance. When radiation density starts to exceed
vacuum density at the last stage of evaporation we would have to change
the equation (16) taking into account the evolution of Hubble parameter
as well as of matter and radiation density, which in the standard
FRW cosmology evolves as $~ t^{-2}$. 
The Eq.(20) reproduces this behaviour starting at the stage of vacuum
dominancy for the case of maximum possible cross-section $\sigma$.
Provided that this behaviour remains dominating at successive stages,
this corresponds to the existence
of remnant evaporating condensate today with the
density comparable to the total density in the Universe which
seems to agree with results of recent analysis 
of observational data \bib{neta}.

The generalization of this approach to the case of arbitrary 
scalar field potential is straightforward. Any cosmologically
reasonable potential must satisfy the condition $V(\phi^2-\phi_0^2)>0$.
True vacuum state $<\phi>=\phi_0$ is determined as the minimum of the
potential $V=0$. The physical particles $\chi=\phi-\phi_0$ are
defined over the true vacuum state. Their mass is given, as usual,
by $\partial^2 V/\partial \phi^2$. Any state with $V(\chi)>0$ we can treat
as bound state of $\chi$ particles trapped inside a Bose condensate.
The equilibrium fluctuations density corresponds to deviation
of the potential from its initial value at the given point
by the quantity $~ \rho_{\chi}+\rho_r~$ (see Eq.(14)). 
It means that in a characteristic time
$m^{-1}$ the field is not completely moved to its ground state,
but, instead, is stabilized near its initial value having slightly
changed by the magnitude $\rho_{\phi}+\rho_r\ll V(\chi)$.

We see that the decoherence of $~ \chi~$ particles and
the back reaction of their decay
products leads to effective freezing of the field near its initial
value. Near this value the potential becomes locally flat, and
the energy density of condensate of $~\chi~$ particles starts to play
a role of an effective cosmological constant. It realizes the case
of chaotic inflation for initially nonflat potential in the case
$m\gg H$ and $~ \phi_0\ll m_{Pl}$. 

At the first sight, the appearance of slow evaporation regime in our
approach seems to lead to the same spectrum of initial density
fluctuations as in slow rolling models. 
However, the origin of fluctuations is different.
In our case fluctuations are generated by statistical distribution
of evaporated particles, while in the typical slow rolling picture
they originate from nonsimultaneous transitions to the ground state.
We are currently investigating this problem and we expect "blue" spectrum
of density fluctuations, which seems to be favored by theories of large
scale structure formation. 

Formally the mechanism presented in this paper is based on rather
trivial solution ($\phi\simeq {const}$) of scalar field dynamics,
which however appears to have nontrivial consequences leading 
to kinetic equilibrium regime for slow evaporation 
of Bose-Einstein condensate. 
This kind of solution has analogies in experimentally studied 
Bose-Einstein condensation in atomic physics \bib{shlap}.

The case of Higgs field considered here is the simple illustration
of the proposed mechanism of $~\Lambda~ $ decay, which seems to be 
more generic and to work also in non-Abelian gauge models 
without Higgs mechanism, in which symmetry breaking is induced 
by nonlinearity of gauge interactions as in technicolor models.

Let us summarize. The kinetics of the Bose
condensate evaporation can effectively damp the coherent field 
oscillations leading 
to slow evaporation regime   
for wide range of possible  particle interaction parameters.
In the cosmological context this 
 provides the effective mechanism for $\Lambda$ decay
responsible for dynamics of symmetry breaking, which can incorporate
inflation, reheating,
as well as nonzero cosmological constant today. 
Currently we are working on rigorous justification  
of the existence of this mechanism
in the frame
of quantum field theory \bib{us00}.
\vskip0.2in
{\bf Acknowledgement}

We are grateful to Yu. M. Kagan, R. Konoplich, A. Sakharov, A. Sudarikov, and
E. Zhizhin for useful discussions. This work was supported by the Polish
Committee for Scientific Research through the grant 2P03D.017.11
and partially performed in the frame of Section "Cosmoparticle
physics" of the Russian State Scientific Technical Programme "Astronomy.
Fundamental Space Research". One of us (M.Kh) expresses his gratitude to
UWM (Poland) and IHES (France) for kind hospitality.\vskip0.2in
{\bf References}
\vskip0.2in
\begin{enumerate}
\item\label{neta}
N. A. Bahcall, J. P. Ostriker, S. Perlmutter, P. J. Steinhardt,
Science {\bf 284}, 1481 (1999) 
\item\label{bel}
M. Bellini, Classical Quantum Gravity {\bf 17}, 145 (2000)
\item\label{berera}
A. Berera and Li-Zhi Fang, Phys. Rev. Lett. {\bf 74}, 1912 (1995)
\item\label{warm}
A. Berera, Phys. Rev. Lett. {\bf 75}, 3218 (1995); Phys. Rev. {\bf D54},
2519 (1996); Phys. Rev. {\bf D55}, 3346 (1997); 
A. Berera, M. Gleiser, R. O. Ramos, Phys. Rev. Lett. {\bf 83}, 264 (1999);
W. Lee and Li-Zhi Fang, Phys. Rev. {\bf D59}, 083503 (1999)
\item\label{boyan}
D. Boyanovsky, H. J. de Vega, R. Holman,
J. F. J. Salgado, in: {\it String theory in curved space times}, 
World Scientific Publishing Company, River Edge, NJ, p. 260 (1998)
\item\label{us96}
S. Capozziello, R. De Ritis, I. Dymnikova, C. Rubano, P. Scudellaro,
Nuovo Cimento {\bf B111}, 623 (1996)
\item\label{spin}
D. Cormier, R. Holman, Phys. Rev. {\bf D60}, 041301 (1999)
\item\label{dolgov}
A. D. Dolgov, in: {\it Proceedings of the Nuffield Workshop
"The very early universe"}, Eds. G. W. Gibbons, S. W. Hawking,
S. T. C. Siklos, Cambridge University Press, p. 449 (1983)
\item\label{dolg}
A. D. Dolgov, Phys. Rev. {\bf D55}, 5881 (1997)
\item\label{me86}
Dymnikova I.G.,  Sov. Phys. JETP, {\bf 63}, 1111 (1986)
\item\label{us95}
I.G. Dymnikova and M. Krawczyk, in: {\it Birth of the Universe}, 
ed. F. Occhionero, Springer Verlag (1995);
Mod. Phys. Lett. {\bf A10}, 3069 (1995)
\item\label{us98}
I. Dymnikova and M. Khlopov,  Gravitation and Cosmology {\bf 4},
50 (1998)
\item\label{us00}
I. Dymnikova, M. Khlopov, A. Sakharov, to be published (2001). 
\item\label{unruh}
R. Fakir and W. G. Unruh, Phys. Rev. {\bf D41}, 1792 (1990)
\item\label{ford}
L. H. Ford,  Phys. Rev. {\bf D31}, 710 (1985)
\item\label{hawking}
G. W. Gibbons and S. W. Hawking,  Phys. Rev. {\bf D15}, 2738 (1977)
\item\label{gliner}
E.B. Gliner,  Sov. Phys. Dokl. {\bf 15}, 559 (1970)
\item\label{us75}
E.B. Gliner, I.G. Dymnikova,  Sov. Astron. Lett. {\bf 1}, 93 (1975)
\item\label{guth}
A. Guth,  Phys. Rev. {\bf D23}, 389 (1981).
\item\label{hiscock}
W. A. Hiscock,  Phys. Lett. {\bf 166B}, 285 (1986)
\item\label{kof}
L. Kofman, A. Linde, A. Starobinski, Phys. Rev. Lett. {\bf 73},
3195 (1994); {\it ibid.} {\bf 76}, 1011 (1996); L. Kofman, in: 
{\it General relativity and relativistic astrophysics}, 
Amer. Inst. Phys., Melville, NY, p.191 (1999)
\item\label{kolb}
E. W. Kolb and M. S. Turner,  The Early Universe, Addison-Wesley (1990)
\item\label{linde}
A. D. Linde, {\it Particle Physics and Inflationary Cosmology},
Harvard Acad. Press, Geneva (1990)
\item\label{mottola}
E. Mottola,  Phys. Rev. {\bf D31}, 754 (1985)
\item\label{myhrvold}
N. P. Myhrvold,  Phys. Rev. {\bf D28}, 2439 (1983)
\item\label{olive}
Keith A. Olive,  Phys. Rep. {\bf 190}, 309 (1990)
\item\label{ostriker}
J. P. Ostriker, P. J. Steinhardt,  Nature {\bf 377}, 600 (1995)
\item\label{cooperstock}
J. M. Overduin a F. I. Cooperstock,  Phys. Rev. {\bf D58}, 
043506 (1998)
\item\label{parker}
L. Parker and Yang Zhang,  Phys. Rev. {\bf D44}, 2421 (1991);
{\bf D47}, 416 (1993)
\item\label{quigg}
C. Quigg, {\it Gauge Theories of the Strong, Weak, and Electromagnetic
Interactions}, Addison-Wesley Publ. Company (1983)
\item\label{shlap}
G. Shlyapnikov et al, in {\it Proc. of the EURESCO Conference
on Bose-Einstein condensation in Atomic Physics}, San Feliu di Guixols, Spain,
September 1999
\item\label{rob}
Y. Shtanov, J. Trashen, R. Brandenberger, Phys. Rev. {\bf D51},
5438 (1995)
\item\label{star}
A. A. Starobinsky,  Phys. Lett. {\bf 91B}, 99 (1980)
\item\label{crit}
J. Yokoyama, A. Linde, Phys. Rev. {\bf D60}, 083509 (1999)
\item\label{zla}
I. Zlatev, G. Huey, P. J. Steinhardt, Phys. Rev. {\bf D57}, 2152 (1998)
\end{enumerate}

\vfill\eject

\end{document}